\documentclass{article}
\usepackage[T1]{fontenc} 
\usepackage[utf8]{inputenc} 
\usepackage{ismir,amsmath,cite,url}
\usepackage{graphicx}
\usepackage{color}
\usepackage[most]{tcolorbox}
\usepackage{lipsum} 
\usepackage{listings} 
\usepackage{lineno}
\usepackage{pdfpages}
\usepackage{graphics}
\usepackage{geometry}
\usepackage{booktabs}  
\usepackage{xcolor}   

\title{ComposerX: Multi-Agent Symbolic Music Composition with LLMs}

\author{
    Qixin Deng$^{5}$, Qikai Yang$^{6}$, Ruibin Yuan$^{1}$, Yipeng Huang$^{2}$, \\ Yi Wang$^{2}$, Xubo Liu$^{8}$, Zeyue Tian$^{1}$, Jiahao Pan$^{1}$, Ge Zhang$^{9}$, Hanfeng Lin$^{2}$, Yizhi Li$^{4}$, Yinghao Ma$^{3}$, \\Jie Fu$^{1}$, Chenghua Lin$^{4}$, Emmanouil Benetos$^{3}$, Wenwu Wang$^{8}$, Guangyu Xia$^{7}$, Wei Xue$^{1}$, Yike Guo$^{1}$
    \\\\
    \textit{$^{1}$Hong Kong University of Science and Technology} \\
    \includegraphics[height=10pt,width=10pt]{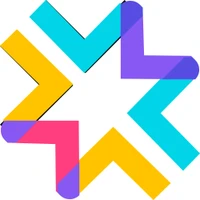} \textit{$^{2}$Multimodal Art Projection Research Community} \\
    \textit{$^{3}$Queen Mary University of London} \\
    \textit{$^{4}$The University of Manchester}\\
    \textit{$^{5}$University of Rochester} \\
    \textit{$^{6}$University of Illinois at Urbana-Champaign}\\
    \textit{$^{7}$Mohamed bin Zayed University of Artificial Intelligence}\\
    \textit{$^{8}$University of Surrey} \textit{$^{9}$01.AI}
}

\begin{document}

\maketitle
\begin{abstract}
Music composition represents the creative side of humanity, and itself is a complex task that requires abilities to understand and generate information with long dependency and harmony constraints. While demonstrating impressive capabilities in STEM subjects, current LLMs easily fail in this task, generating ill-written music even when equipped with modern techniques like In-Context-Learning and Chain-of-Thoughts. To further explore and enhance LLMs' potential in music composition by leveraging their reasoning ability and the large knowledge base in music history and theory, we propose \textbf{ComposerX}\footnote{Demo page: https://lllindsey0615.github.io/ComposerX\_demo/}, an agent-based symbolic music generation framework. We find that applying a multi-agent approach significantly improves the music composition quality of GPT-4. The results demonstrate that ComposerX is capable of producing coherent polyphonic music compositions with captivating melodies, while adhering to user instructions.
\end{abstract}
\section{Introduction}
\label{sec:intro}
Music shares many structural similarities with language~\cite{masataka2009origins, masataka2007music, pino2023association}, prompting researchers to explore the application of language models (LMs) in music generation~\cite{vaswani2017attention,huang2018musictransformer,payne2019musenet,lu2023musecoco,dhariwal2020jukebox,agostinelli2023musiclm,copet2023simple,margulis2016repetition,dai2022missing,jhamtani2019modeling,qu2024mupt}. Recent advances in large language models (LLMs) have opened potential pathways towards achieving Artificial General Intelligence (AGI). While much of the research emphasis has been on the STEM aspects of AGI~\cite{yue2023mammoth,roziere2023code,bubeck2023sparks}, there is comparatively less focus on its creative dimensions, particularly in music creation. Current methodologies primarily involve training LMs from scratch, as seen with initiatives like MusicLM~\cite{agostinelli2023musiclm} and MusicGen~\cite{copet2023simple}, with a predominant focus on audio generation. However, these models often struggle with processing advanced musical instructions and typically offer only limited control options, such as genre and instrument selection. Enhancing controllability in these systems requires neural architectural engineering and extensive computational resources~\cite{lin2024arrange,lin2023content,lin2023equipping}.

Recent research, influenced by Bubeck et al.~\cite{bubeck2023sparks}, has revealed that pretrained large language models (LLMs) might inherently possess emergent musical capabilities. Inspired by these findings, subsequent studies~\cite{yuan2024chatmusician,ding2024songcomposer,liang2024bytecomposer} have explored leveraging pretrained LLM checkpoints for handling symbolic music in an end-to-end manner, aiming to tap into the extensive knowledge and reasoning abilities embedded in these LLMs. 
However, these unified approaches are not without limitations. They depend heavily on hand-crafted datasets tailored for specific musical tasks and often require both a phase of continual pretraining and subsequent supervised fine-tuning. Furthermore, while training on symbolic music data is generally less computationally intensive than processing raw audio data, the costs remain prohibitive for many researchers. For example, renting an 8xGPU machine (such as a p4d.24xlarge spot instance on AWS) for one month can exceed \$8,000 USD\footnote{https://instances.vantage.sh/aws/ec2/p4d.24xlarge}, posing a significant financial barrier.

In this paper, we introduce a novel multi-agent-based methodology, ComposerX\footnote{https://github.com/lllindsey0615/ComposerX}, which is training-free, cheap, and unified. Leveraging the internal musical capabilities of the state-of-the-art GPT-4-turbo, ComposerX can generate polyphonic music pieces of comparable, if not superior, quality to those produced by dedicated symbolic music generation systems\cite{lu2023musecoco,wu2023exploring} that require extensive computational resources and data. ComposerX utilizes approximately 26k tokens per song, incurring a cost of less than \$0.8 USD per piece. Throughout the development phase of ComposerX, the total expenditure on the OpenAI API was under \$1k USD. We achieved a good case rate of 18.4\%, as assessed by music experts, which translates to an average cost of approximately \$4.34 USD for each musically interesting piece. Furthermore, experimental results demonstrate that the multi-agent strategy substantially enhances composition quality over single-agent baselines. In Turing tests, approximately 32.2\% of the pieces identified as `good' by ComposerX were indistinguishable from those composed by humans, as indicated in Table \ref{tab:turing}.

While there is existing research on musical LLM agents~\cite{zhang2023loop,yu2023musicagent}, our approach distinctively diverges from these precedents. Prior studies primarily focus on single-agent systems. In contrast, our work introduces a multi-agent framework, emphasizing collaborative aspects of music creation. Furthermore, we concentrate on symbolic music generation, leveraging the intrinsic musical understanding of LLMs without the need for external computational resources or tools. Previous methodologies typically depend on GPU servers for deploying local inference services, treating the LLMs more as tool-use agents rather than harnessing their inherent capabilities to process and generate musical content. In sum, the contributions of our paper are as follows:

(1) We propose the first multi-agent polyphonic symbolic music composition system, ComposerX. It elicits the internal musical capabilities inside LLMs without the need for external tools.

(2) Through extensive subjective evaluations, we demonstrate that our multi-agent approach substantially enhances the quality of music composition compared to single-agent systems and specialized music generation models. Our method also offers cost-efficiency advantages by obviating the need for dedicated training or local inference services.

(3) We commit to the advancement of this research area by open-sourcing our code, prompt-set, and experimental results, facilitating further investigation and development by the community.

\section{Method}
We first construct a set of user prompts for music composition, which is used for evaluation. Then we demonstrate how we implement our single-agent and multi-agent LLM composition systems.
\subsection{User Prompt Set Curation}
To understand how the users, typically those with substantial musical backgrounds, would prompt a text-to-music generation system, a user prompt set is collected by asking humans with music backgrounds to manually write high-quality prompts. These prompts typically include essential musical attributes such as genre, tempo, key, chord progression, melody, rhythm, number of bars, number of voices, instruments, style, feeling, emotion, title, and motif of the music piece. Based on the human-written samples, more prompt samples are generated using Self-instruct by GPT-4\cite{wang2023selfinstruct}. This results in a set of 163 prompts, which is used in the later agent testing and system evaluation. 
An example prompt is given below. 
\begin{tcolorbox}[title=Prompt, fonttitle=\bfseries\small, coltitle=black, colbacktitle=gray!20, colback=white, colframe=gray!50, boxsep=1mm, left=1mm, right=1mm, top=1mm, bottom=1mm, fontupper=\small]
    \textbf{Vintage French Chanson:} A nostalgic chanson in C major with a slow tempo, featuring accordion, violin, and upright bass over 16 bars with chords C, Am, Dm, G. The accordion leads with expressive sound, violin adds romance, and the upright bass supports, evoking vintage French charm.
\end{tcolorbox}
\begin{tcolorbox}[title=Attributes, fonttitle=\bfseries\small, coltitle=black, colbacktitle=gray!20, colback=white, colframe=gray!50, boxsep=1mm, left=1mm, right=1mm, top=1mm, bottom=1mm, fontupper=\small]
\textbf{Name:}  Vintage French Chanson\ 
\textbf{Tempo:} Slow\\ \textbf{Feeling:} Nostalgic\ \textbf{Chord Progression:}  C, Am, Dm, G\\
\textbf{Key:}  C major \ \
\textbf{Bars:}  16 \ \ 
\textbf{Instruments:}  Accordion, violin, upright bass 
\end{tcolorbox}

\subsection{Single-Agent}
We apply various prompt engineering techniques, including In Context Learning (ICL), Chain of Thought (CoT), and Role-play to guide a single GPT acting as the composer. Additionally, we have refined the prompt template by incorporating specific instructions that ensure the correctness of the ABC notation format.

\textbf {Original GPT with Simple Role-play (Ori):} To investigate the inherent capabilities of the original GPT model in interpreting user prompts and generating ABC notation, we instructed GPT to act in the role of a professional composer, with user prompts directly input into the system. This method aims to assess the model's basic performance in music composition without the integration of additional complex prompting techniques.

\textbf {Role-Play with Additional Instruction (Role):} Inspired by classical rule-based computer music generation, we equipped GPT with enhanced musical knowledge focusing on phrase management and melody line construction, detailed in Table \ref{tab:single_prompt_role}. For instance, in composing melodies, we instructed the model to ensure that each generated melody possesses distinct phrase divisions. Additionally, each phrase is required to conclude with a prominent ending note, which serves as the last note of the phrase. By incorporating these additional instructions, we aim to elevate the music quality and structural coherence of the music, aligning the generated compositions more closely with traditional musical standards.
\begin{table}[!ht]
\centering
\label{tab:single_prompt_role}
\scalebox{0.95}{ 
\begin{tabular}{|p{0.95\linewidth}|}
\hline
\textbf{\small Role-play Prompting with Additional Music Knowledge} \\ 
\hline
\ttfamily
\small{
{\color{blue}You are a talented musician.} 
 Here are some tips for generating melodies: 
 
1. The generated melody should have clear phrase divisions, and it's preferable to avoid more than two consecutive measures within one phrase to prevent an uncomfortable listening experience. There should be a certain amount of space between phrases, allowing the audience to clearly distinguish between them.  

2. A phrase usually has a prominent ending note, which is the last note of the entire phrase. It typically has a longer duration, or it might be followed by a rest. This ending note is usually within the key or the chord, e.g., phrases ending with a Cmaj chord usually terminate on one of the three chord tones, C, E, or G, ensuring a stable listening experience.

3. When generating melodies, the movement of the notes should primarily consist of stable intervals such as whole steps, thirds, and fifths, while avoiding excessive large leaps. This will help maintain a sense of logic and coherence throughout the composition. 

4. The rhythm of the phrases should be rich and harmonious. Try using different rhythmic patterns to build the melody, such as combining eighth notes with sixteenth notes, syncopated rhythms, or triplets.}

\\
\hline
\end{tabular}
}
\caption{Single-agent role-play(indicated in the blue text) prompting with additional tips given by human composer on melody construction.}
\label{tab:single_prompt_role}
\vspace{-4mm}
\end{table}

\textbf {Chain-of-Thought (CoT):} As proven in other fields of research, CoT improves the ability of LLMs on complex reasoning by encouraging them to write down intermediate reasoning steps\cite{wei2023chainofthought}. Within the context of music composition, we deconstruct the music generation process into several distinct stages. These stages include specifying initial music information, such as title, key, tempo, and speed, followed by the development of chord progressions and melody composition as detailed in Table \ref{tab:single_prompt_cot}. 
\begin{table}[ht!]
\centering
\scalebox{0.95}{ 
\begin{tabular}{|p{0.95\linewidth}|}
\hline
\textbf{\small Chain of Thought prompting with three steps} \\ 
\hline
\ttfamily
\small{
First, you need to determine all the information related to the piece in the ABC notation format, such as the name,tune, speed, mode, and anything other than the notes. 
This forms the basis of the piece's style.***Note that only return the music information in ABC notation format without any notes or text or Additional note.*** 

Second,Based on the song information in the ABC notation format provided earlier, generate a ***16-bar long*** chord progression and return it in text form, with each bar separated by a "|" symbol. The generated chord progression should be consistent with the song's key and as closely aligned with the song's theme and characteristics as possible.

Now the chord progression and other information are provided,you are required to create a ***16-bar long*** piece of music based on these information.}

\\
\hline
\end{tabular}
}
\caption{Single-agent CoT prompting method with three steps.}
\label{tab:single_prompt_cot}
\vspace{-4mm}
\end{table}

\textbf {In Context Learning (ICL):} ICL leverages a few input-output examples to enhance an LLM’s understanding of a specific task. In this method, we use pairs of user prompts and corresponding ABC notations from ChatMusician\cite{yuan2024chatmusician} as demonstrative examples for prompting as detailed in Table \ref{tab:single_prompt_icl}. 
\begin{table}[ht!]
\centering
\scalebox{0.95}{ 
\begin{tabular}{|p{0.95\linewidth}|}
\hline
\textbf{\small Single-agent In-context learning prompting method} \\ 
\hline
\ttfamily
\small{
You are an intelligent agent with musical intelligence, and your goal is to create music that meets the relevant needs and human listening habits.In this task, use ABC as the format for outputting sheet music.***Only return the ABC notation without any other description or text,and only return one piece that follow the music description given this time.***Below are the requirements for the music,it contains music elements like title,genre,key and more,and some composition examples are listed after the requirements.

{\color{red} ABC Notation of the selected sample}
}

\\
\hline
\end{tabular}
}
\caption{Single-agent In-context learning prompting method}
\label{tab:single_prompt_icl}
\vspace{-4mm}
\end{table}

\subsection{Multi-Agent Music Composition: ComposerX}
To enhance the music generation capabilities of GPT-4, we developed a collaborative music creation framework, ComposerX, that draws inspiration from key elements inherent in real-world music composition processes, such as melody construction, harmony or counterpoint development, and instrumentation. This framework facilitates the music creation process through a structured conversation chain between agents role-played by GPT-4.
\subsubsection{Agent Role Assignment}
In the collaborative music creation framework designed to augment GPT-4's music generation capabilities, roles are assigned to ensure a structured and efficient composition process. The assignment of roles is as follows:

\noindent\textbf{Group Leader}: Tasked with interpreting user inputs, decomposing these inputs into granular tasks, and assigning these tasks to specialized agents in the group. 

\noindent\textbf{Melody Agent}: Responsible for generating single-line melodies under the guidance of the group leader.

\noindent\textbf{Harmony Agent:} This agent is tasked with enriching the musical piece, and adds harmonic and contrapuntal elements to the melody.

\noindent\textbf{Instrument Agent:}This agent selects and assigns instruments to each voice.

\noindent\textbf{Reviewer Agent:}Performing a quality assurance role, this agent evaluates the outputs of the melody, harmony, and instrumentation agents across four critical dimensions. (1)Melodic Structure: Evaluation of melody's narrative flow, thematic development, and variation in pitch and rhythm.
Harmony and Counterpoint: Assessment of how harmonies complement the melody, counterpoint effectiveness, and chord progression quality. (2)Rhythmic Complexity: Analysis of rhythm's role in sustaining interest, its synergy with melody, and the incorporation of dynamic variations. (3)Instrumentation and Timbre: Review of instrument selection, timbral blending, and dynamic usage to achieve an optimal auditory experience. (4)Form and Structure: Examination of the composition’s overarching structure, transitional elements, connectivity between sections, and conclusion efficacy.

\noindent\textbf{Arrangement Agent:} Concluding the collaborative process, this agent is responsible for compiling and formatting the collective output into standardized ABC notation, ensuring the music is documented in a universally readable format.

\subsubsection{Agent Communication Pattern}
The collaborative framework employs a structured communication pattern to ensure an orderly and efficient flow of information between the different agents involved in the composition process. This pattern is crucial for maintaining the integrity and coherence of the musical piece being generated. The communication process unfolds as follows:

\noindent\textbf{Initial Composition Round}: The composition process begins with the Group Leader Agent initiating the sequence by analyzing the user input and breaking it down into specific tasks assigned to the Melody, Harmony, and Instrument Agents respectively. This step sets the foundation for the composition based on the user's requirements.
Following the leader's instructions, the Melody Agent then generates the initial melody line, adhering to the thematic direction and stylistic guidelines provided by the Group Leader.
Subsequently, the Harmony Agent enriches the melody by adding harmonic layers and counterpoints.
The Instrument Agent assigns appropriate instruments to the generated melody and harmony lines by selecting timbres that complement the overall composition.

\textbf{Iterative Review and Feedback Cycle}: Upon completion of the initial composition round, the Reviewer Agent steps in to evaluate the work produced by the Melody, Harmony, and Instrument Agents. This agent provides comprehensive feedback across several critical dimensions, including melodic structure, harmony and counterpoint, rhythmic complexity, and instrumentation.

Based on the feedback from the Reviewer Agent, the Melody, Harmony, and Instrument Agents proceed to refine their respective components of the composition. This iterative refinement process typically follows the order: Melody, Harmony, and then Instrument, allowing for modifications to be made in response to the feedback provided.

The composition undergoes several rounds of review and refinement, with the Reviewer Agent continuously providing feedback to ensure the musical piece evolves toward a coherent and high-quality final product. This iterative process allows for dynamic adjustments and enhancements to be made, enriching the overall composition.

\textbf{Final Arrangement and Notation}: Once the composition has reached a satisfactory level of polish and coherence, the Arrangement Agent takes over to compile and format the collective output into the standardized ABC notation. This final step ensures that the music is documented in a format that is readable and can be interpreted by musicians and software alike.

This communication pattern enables a collaborative and adaptive approach to music generation. The primary advantage of this communication pattern lies in its ability to simulate a real-world collaborative music creation environment, where each participant's input is valued and considered in the development of the final piece. Integrating this structured approach with a multi-agent system significantly reduces the likelihood of LLM hallucination, a notable challenge where models may generate inaccurate or nonsensical outputs. By assigning specific roles to specialized agents, the framework ensures that each segment of the music composition undergoes rigorous scrutiny and refinement. This division of labor not only enhances the precision and relevance of the generated content but also leverages the collective intelligence of the agents to cross-verify information, thereby mitigating the risk of incorporating erroneous elements into the composition. An example of the music composition process is given in Figure 1 and Figure 2 with a comparison made between

\begin{figure*}[!t]
\centering
\includegraphics[width=.8\textwidth]{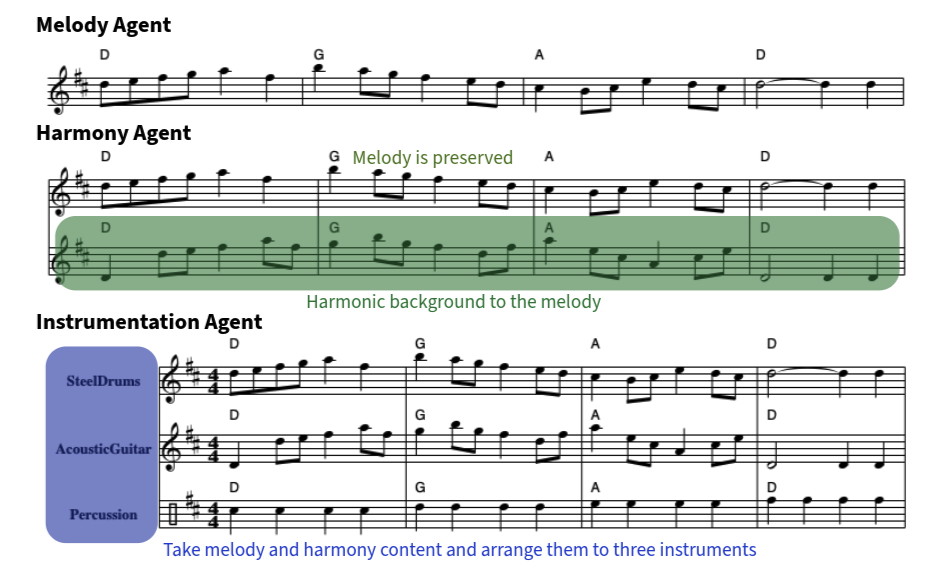}
\caption{The Leader Agent will distribute the tasks among the Melody Agent, Harmony Agent, Instrumentation Agent when it is requested a "Breezy Caribbean Calypso" piece. This figure demonstrates the work of the three agents with changes in the same four bar opening.}
\label{fig:leader_agent}

\includegraphics[width=.8\textwidth]{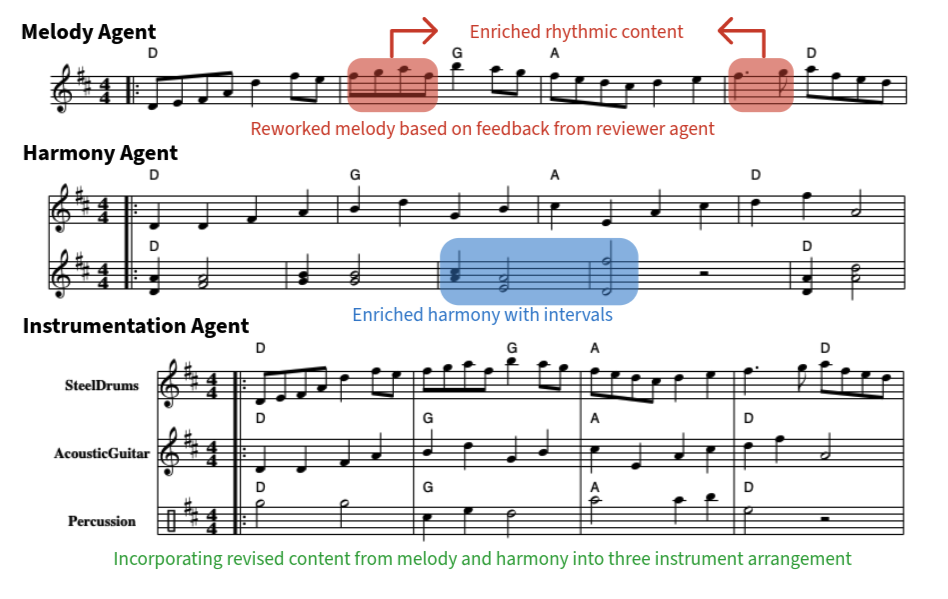}
\caption{The Reviewer Agent then analyze the collective effort of the three agents in the first stage (shown in Figure \ref{fig:leader_agent}), and give advice for agents to work on. This figure demonstrates the work of the three agents after incorporating the advice given by the Reviewer Agent in the same four-bar opening.} 
\label{fig:reviewer_agent}
\end{figure*}

\begin{figure*}[ht] 
    \centering
    \includegraphics[width=0.9\textwidth]{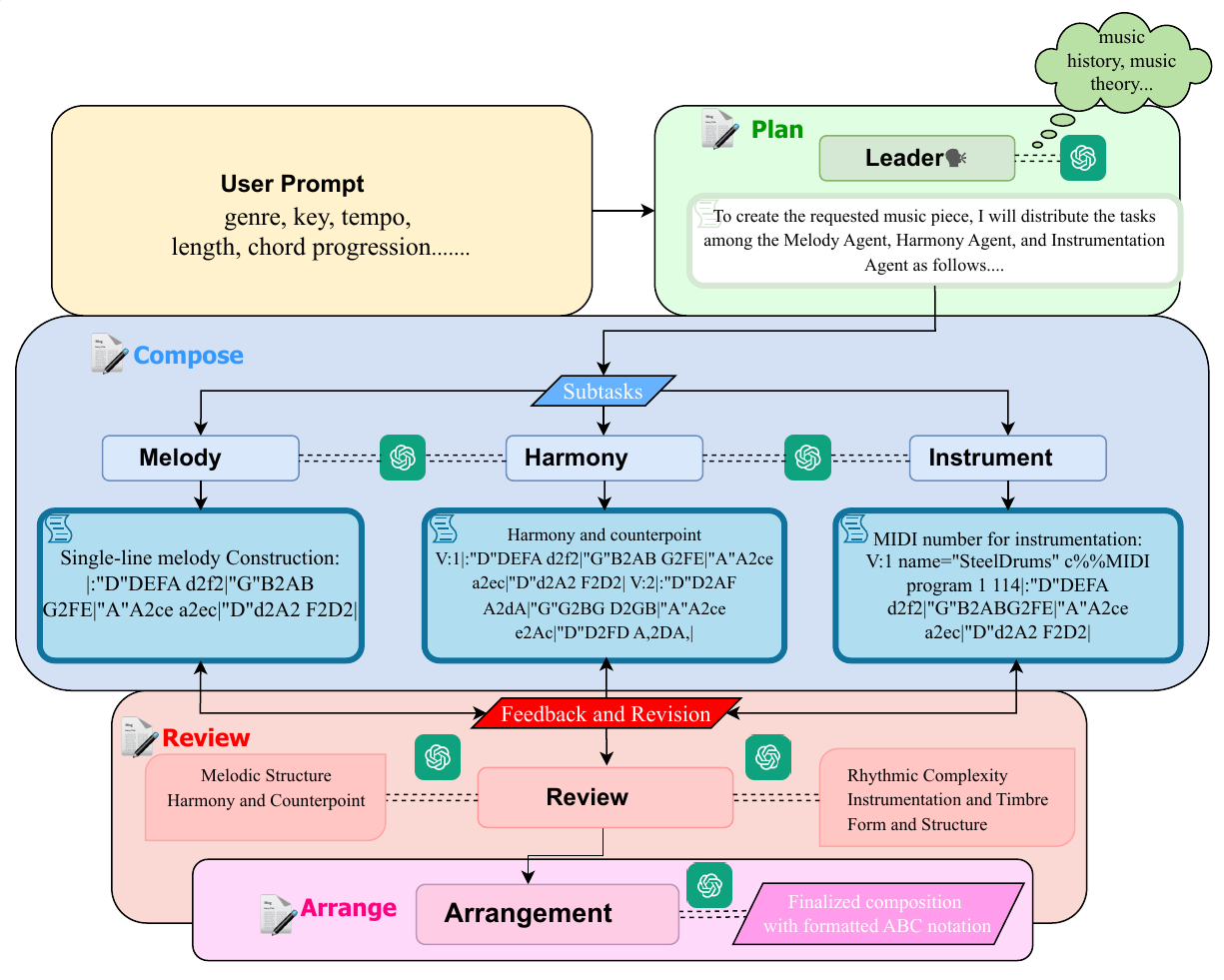} 
    \caption{Agent Communication Pattern of ComposerX.The system is given with a user prompt. In the Planning stage, the Leader analyzes the user prompt and decomposes it into subtasks that can be assigned to other musician agents. In the Composing stage, the musician agents, including Melody Agent, Harmony Agent, and Instrument Agent compose in ABC notation according to their assigned tasks. In the Reviewing stage, the Review Agent provides constructive feedback to the musician agents and the musician agents revise their work according to the feedback they received. In the arrangement stage, the Arrangement Agent arranges the work of the musicians agent to standardized ABC notation.}
    \label{composition}
\end{figure*}

\subsubsection{Agent Prompt Engineering}
Agent prompt engineering emerges as a crucial technique for optimizing the performance of each specialized agent and the quality of the generated music. This process involves the meticulous design of role-specific instructions and guidelines that encapsulate both the musicality and technicality of ABC notation generation. The framework incorporates In-Context Learning for ABC notations to ensure agents can effectively communicate and document their contributions. An example of the agent prompt is given in Table \ref{tab:multi_prompt}. This section elaborates on these components and their significance in fostering collaborative dynamics within the framework.

\textbf{Role-Specific Instructions}: Within the framework, each agent is endowed with a set of instructions tailored to its designated role. These instructions serve to ensure a comprehensive understanding of the agent's duties, the expectations for its performance, and its role within the larger collaborative ensemble. Agents are briefed on the specific outcomes they are expected to achieve and informed about the dynamics of their interactions with other agents. This detailed prompt design facilitates a cohesive operation among the agents, fostering an environment where each component of the framework is aligned toward the collective goal of generating sophisticated and coherent musical compositions.

\textbf{In-Context Learning for ABC Notation}: In Context Learning for ABC notation ensures accurate format output from each agent. The Melody Agent is shown with an example of a monophonic melody in ABC notation, providing a clear model for representing single-line melodies.
The Harmony Agent receives a polyphonic music piece example in ABC notation, aiding in understanding the notation of harmonies and counterpoints in multiple voices.
The Instrument Agent is given a polyphonic piece with MIDI program information noted, demonstrating how to detail instrumental assignments within the notation.
This approach equips agents with the knowledge to correctly apply ABC notation, essential for the structured and coherent documentation of musical compositions.

\begin{table}[tb!]
\centering
\scalebox{0.95}{
\begin{tabular}{|p{0.95\linewidth}|}
\hline
\small{\textbf{Melody Agent Prompt}}\\
\hline
\ttfamily
\small{
{\color{blue}You are a skillful musician, especially in writing melody. }

You will compose a single-line melody based on the client's request
and assigned tasks from the Leader.

You must output your work in ABC Notations.

{\color{red}
Here is a template of a music piece in ABC notation, in this template:

\ \ X:1 is the reference number. You can increment this for each new tune.

\ \ T:Title is where you'll put the title of your tune.

\ \ C:Composer is where you'll put the composer's name.

\ \ M:4/4 sets the meter to 4/4 time, but you can change this as needed.

\ \ L:1/8 sets the default note length to eighth notes.

\ \ K:C sets the key to C Major. Change this to match your desired key.

The music notation follows, with |: and :| denoting the beginning
and end of repeated sections.

Markdown your work using ```    ``` to the client.

```

X:1

T:Title

C:Composer

M:Meter

L:Unit note length

K:Key

|:GABc d2e2|f2d2 e4|g4 f2e2|d6 z2:|

|:c2A2 B2G2|A2F2 G4|E2c2 D2B,2|C6 z2:|

```
}

You will output the melody following this template, 

but decide the time signature, key signature, and the

actual musical contents and length yourself.

After you receive the feedback from the Reviewer Agent,

please improve your work according to the suggestions you were given.

}\\
\hline
\end{tabular}
}
\caption{Prompt for Melody Agent. GPT is prompted with role-specific instructions(indicated in blue text) and In-Context-Learning of ABC notations(indicated in red text)}
\vspace{-4mm}
\label{tab:multi_prompt}
\end{table}

\section{Experiments}

\subsection{Setup}
Our experiment leverages the multi-agent conversation provided by the AutoGen framework\cite{wu2023autogen}, utilizing its group chat function to facilitate a customized interaction among pre-defined agents. This setup comprises an ensemble of agents including one leader, three musician agents (melody, harmony, and instrument agents), one review agent, and one arrangement agent. Additionally, a user proxy agent is integrated into the framework to simulate user interaction by inputting prompts from our curated user prompt set.

We employ the "GroupChatManager" class from AutoGen to ensure seamless coordination and oversight of the conversation's content and workflow. According to AutoGen, the group manager is also powered by LLMs, functions as the supervisor of the conversation, and implements a structured communication protocol that involves three critical steps: dynamically selecting a speaker from the agents, collecting the response from the chosen agent, and disseminating the collected response to the rest of the group. 

For the purpose of our experiment, we have predetermined a maximum number of twelve rounds for agent communication. This limitation allows us to observe the effectiveness of the multi-agent system over a defined number of interaction cycles, facilitating one or more rounds of iterative review and refinement within the conversation. This structured experimental design is aimed at evaluating the collaborative dynamics and output quality of the multi-agent conversation in generating cohesive and musically rich compositions based on user prompts.

\subsection{Evaluation}
\subsubsection{Automatic Evaluation}

We conducted two experiments to automatically evaluate our system. One experiment assessed the success rate of generating symbolic music in a multi-agent setting, with results presented in Table~\ref{tab:success_rate}. One experiment compared the sequence lengths of symbolic music generated by multi-agent and single-agent systems, detailed in Table~\ref{tab:string_len}. These experiments demonstrate the effectiveness of our approach in generating symbolic music.

\begin{table}[h]
\centering
\small
\begin{tabular}{lccc}
\toprule
    Checkpoints           & Generation Success Rate \\
\midrule
GPT-4-Turbo & $98.2$\%  \\

GPT-4-0314 & $95.7$\%  \\

GPT-3.5-Turbo & $73.0$\%  \\
\bottomrule
\end{tabular}
\caption{\textcolor{black}{One-time generation success rate for multi-agent system with different checkpoints}}
\label{tab:success_rate}
\end{table}

\begin{table}[h]
\centering
\small
\begin{tabular}{lccc}
\toprule
    Methods           & Average ABC String Length \\
\midrule
GPT-4-Turbo multi & $1005.925$  \\

GPT-4-Turbo cot & $360.92$  \\

GPT-4-Turbo icl & $366.30$  \\

GPT-4-Turbo ori & $354.53$  \\

GPT-4-Turbo role & $337.64$  \\
\bottomrule
\end{tabular}
\caption{\textcolor{black}{The average length of ABC String generated by different methods on GPT-4-Turbo checkpoint}}
\label{tab:string_len}
\end{table}

\subsubsection{Human Listening Test}

To qualitatively assess our work, we conducted three listening tests. In the first test, we compared music samples generated by single-agent and multi-agent baselines. Similar to the AB-test setting from previous work\cite{donahue2023singsong,yuan2024chatmusician}, participants were presented with pairs of samples: one from a multi-agent baseline with GPT-4 Turbo checkpoints, and the other from a single-agent baseline employing prompting techniques mentioned above: Original, In-Context Learning (ICL), Chain of Thought (CoT), and Role-play, also driven by GPT-4 Turbo checkpoints. Participants were asked to select the sample they preferred. All paired samples were generated using the same prompt; however, participants were not informed about the specific prompt details before making their selections.

\begin{figure}[ht!]
    \centering
    \includegraphics[width=0.5\textwidth]{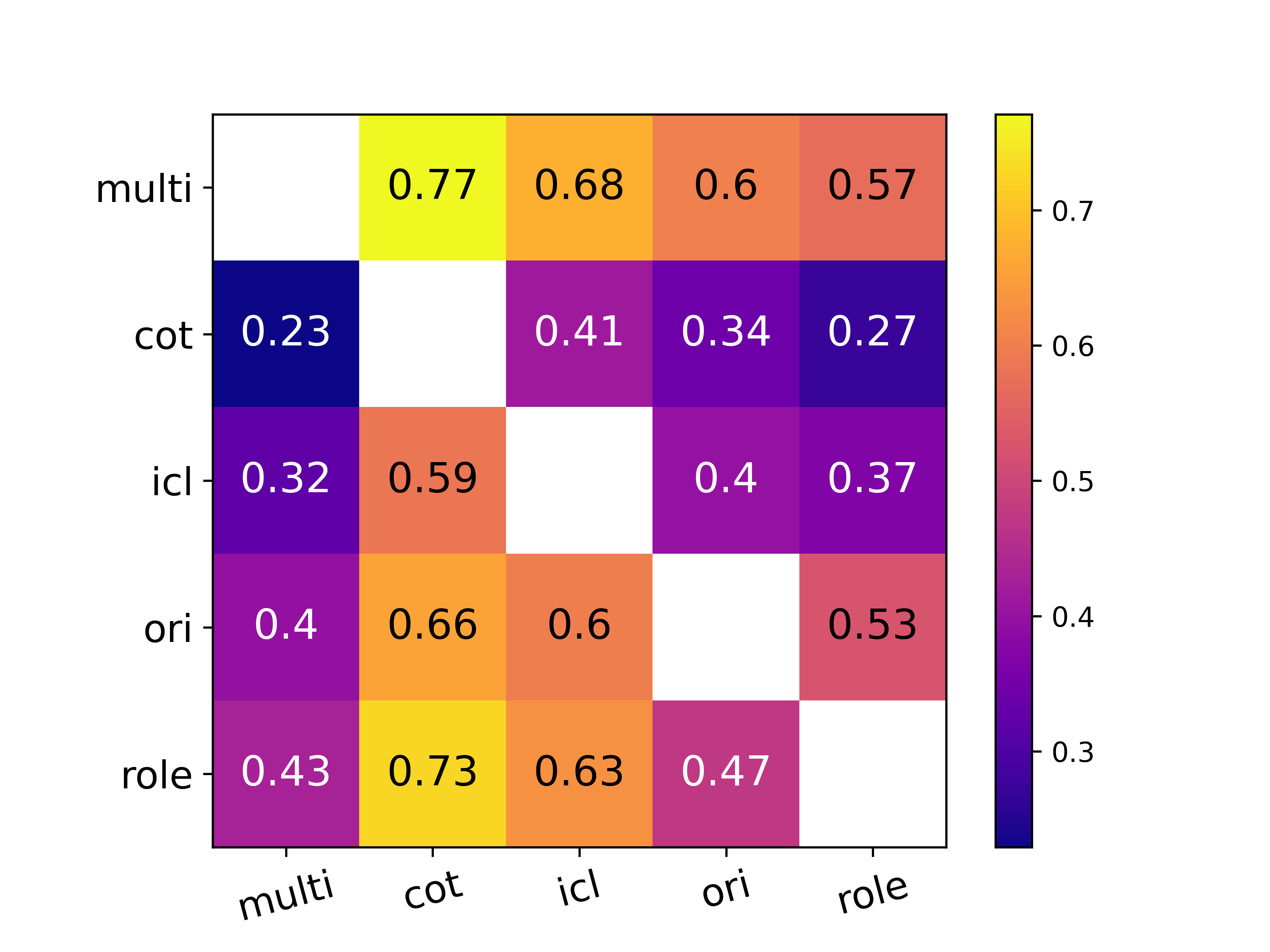}
    \caption{Result from the first listening test comparing multi-agent baseline and single-agent baselines with different prompting techniques. Each row indicates the fraction of listeners' preference for the indicated baseline over other baselines. i.e. 0.77 means raters prefer multi-agent system over CoT single-agent 77\% of the times. }
    \label{fig:exp1_results}
\end{figure}

In the second listening test, we assess the perceived human-like quality of music generated by the multi-agent baselines. Participants were presented with two types of music samples: those generated by multi-agent baselines and those composed by humans, sourced from Irishman and KernScores\footnote{http://kern.ccarh.org/}, which are ABC notation datasets containing human-composed music pieces from all around the world. Each participant is asked to determine whether each sample was composed by a human or a machine.

In the third listening test, we assessed the performance of our multi-agent baselines, which incorporate GPT-4 Turbo, GPT-4-0314, and GPT-3.5-Turbo checkpoints, against established text-to-music generation models. Specifically, comparisons were made with MuseCoco\cite{lu2023musecoco}, developed by Peiling Lu et al., and a BART-based model fine-tuned on 282,870 English text2music pairs in ABC notation, as proposed by Wu et al\cite{wu2023exploring}. Participants were presented with music samples alongside their corresponding prompts and asked to select the sample that best matched the prompt in terms of musical structure and content.
\begin{figure}
    \centering
    \includegraphics[width=0.5\textwidth]{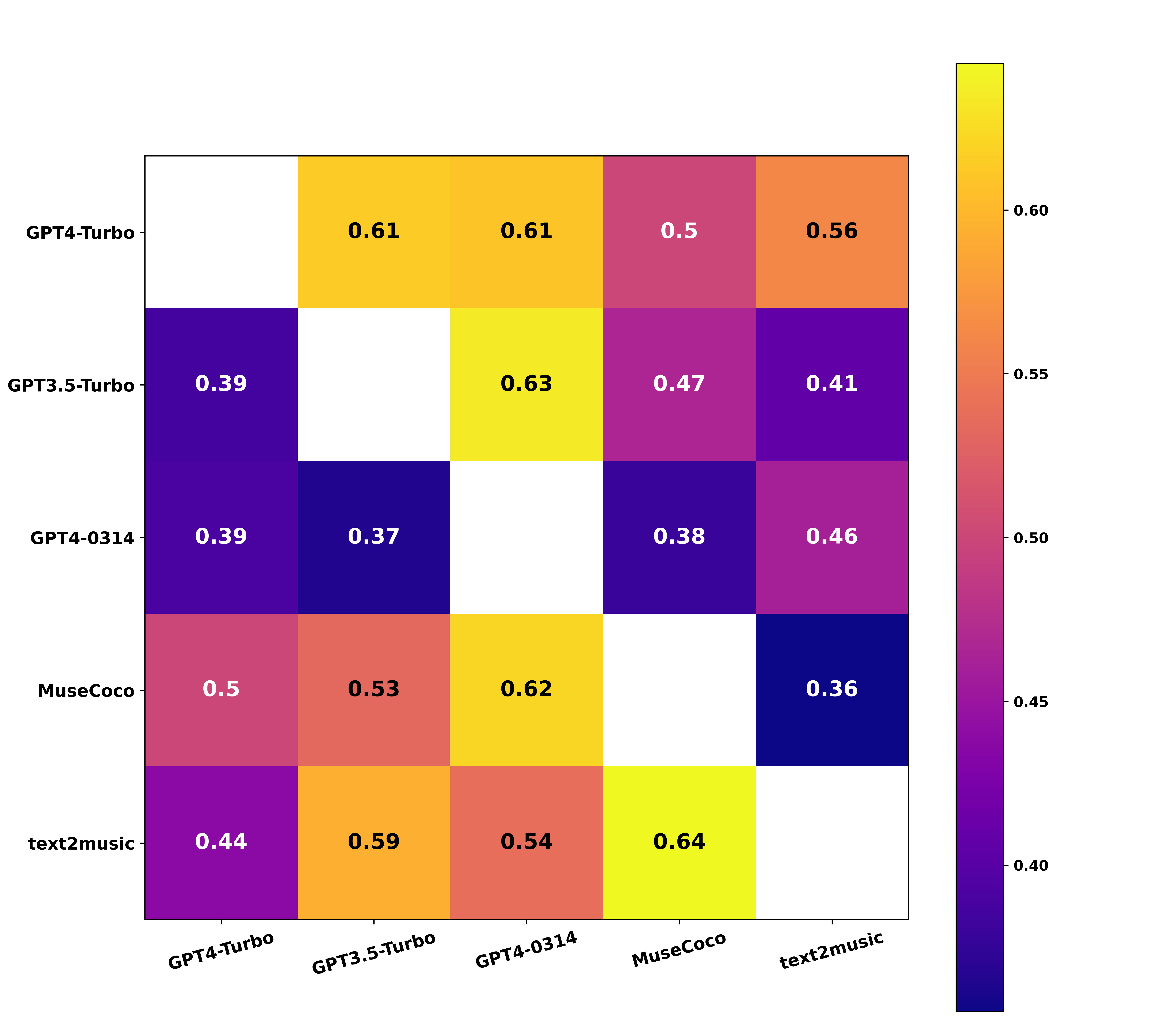}
    \caption{Result from listening test comparing multi-agent baselines with GPT-4-Turbo, GPT-4-0314, GPT-3.5-Turbo checkpoints, MuseCoco and text2music Baselines. Each row indicates the fraction of listeners' preference for the indicated baseline over other baselines. In this case, the strongest multi-agent baseline with GPT-4-Turbo checkpoints outperformed text2music, and received the same score as MuseCoco. }
    \label{fig:exp3_results}
\end{figure}
\subsection{Results}
Results from comparing multi-agent baseline and single-agent baseline appear in Figure 4. The preference score of GPT-4-Turbo multi has 0.77, 0.68, 0.6, and 0.57 on each of other single-agent baselines. 

As indicated by the fractions, the multi-agent baseline outperformed all the single-agent baselines. In addition to the dominant performance of the multi-agent baseline over single-agent baselines, we also observe that the multi-agent baseline has a stronger ability to generate longer music. As indicated by Table \ref{tab:string_len}, GPT-4-Turbo multi demonstrates the capacity to generate music pieces nearly three times longer than those produced by any other single-agent baselines.

\begin{table}[h]
\centering
\small
\begin{tabular}{lccc}
\toprule
    Model           & Perceived as Human &Perceived as Machine  \\
\midrule
ComposerX & $32.2$\% & $67.8$\%  \\
Ground Truth & $55.4$\% & $44.6$\% \\
\bottomrule
\end{tabular}
\caption{\textcolor{black}{Result from our second listening test (Turing test).}}
\label{tab:turing}
\end{table}

Results from comparing the multi-agent baseline with music composed by humans indicates that ComposerX gets 32.2\% perceived as human which is lower than the rate of real human music - 55.4\% as indicated in Table \ref{tab:turing}. Despite failing the Turing test, ComposerX showcases its capability to closely match human skill in music composition.

Results from comparing the multi-agent baseline with GPT-4-Turbo, GPT-4-0314, GPT-3.5-Turbo checkpoints, MuseCoco, and text2music are presented in Figure 5. As indicated by the fractional numbers, the multi-agent baseline with GPT-4-Turbo checkpoints is our strongest-performed baseline. It outperformed text2music baseline with 0.56 preference score and received the same score as MuseCoco. GPT-4-Turbo also shows the highest generation success rate among all checkpoints, as indicated in Table \ref{tab:success_rate}.

\section{Discussion}
Overall, we observed that our GPT-powered multi-agent framework significantly enhances the quality of the music generated over solutions utilizing a singular GPT instance. Advantages of our system include:

\textbf{Controllability:} Our observations of the collaborative interactions among the agents, with particular emphasis on the contributions of the Group Leader, indicate the system's competency in comprehending and executing a wide range of musical attributes as delineated by user inputs. Fundamental components such as tempo, key, time signature, chord progression, and instrumentation are adeptly translated into the corresponding ABC notations. This adept interpretation and realization of user directives significantly augment user controllability, enabling the generation of music that closely mirrors their specifications and artistic preferences.

\textbf{Training-free and data-free:} In contrast to conventional text-to-music generation models, which predominantly depend on large datasets for training, our system introduces significant benefits by obviating the need for such extensive data. This approach substantially mitigates the challenges associated with the compilation and refinement of large training datasets, including potential biases and the substantial resources often requisite for these processes. Moreover, this strategy enhances the system's adaptability and accessibility, marking a shift towards more resource-efficient practices in music generation. As a result, this innovation plays a pivotal role in democratizing music generation, making it more attainable across a broader spectrum of users and applications.

The system exhibits certain limitations, particularly when engaging with the nuanced aspects of musical composition that are often intrinsic to human-created music. These limitations delineate areas for potential enhancement and further research:

\textbf{Subtlety in Musical Expression:} The system is adept at interpreting basic musical elements but faces challenges in generating compositions with the nuanced subtlety characteristic of human composers. This includes aspects such as emotional depth, dynamic contrasts, and intricate phrasing, which are essential for conveying more profound musical narratives and experiences.

\textbf{Translation from Natural Language to Musical Notation:} Instructions and feedback given by the Group Leader and Review Agent aiming to facilitate nuanced musical elements are sometimes inadequately translated into ABC notations by the musician agents. This gap between conceptual understanding and practical embodiment in music notation underscores the system's current limitations in realizing more sophisticated musical ideas.

\textbf{ Instrumental Note Range Compliance:} The system occasionally generates notes beyond the conventional pitch ranges of certain instruments. For example, despite directives to adhere to instrument-specific pitch ranges, outputs have included notes exceeding the upper limit of a contrabass (C2 to F4), reflecting a discrepancy between the system's outputs and practical musical performance constraints.

\textbf{Inter-Voice Alignment:} Our system faces challenges with aligning multiple musical voices accurately. This challenge primarily arises from the inherent limitations of text-based LLMs in generating polyphonic ABC notations. The linear nature of text-based input and output mechanisms does not naturally accommodate the complexity of polyphonic music, where multiple voices or instruments must be coordinated in time. 

\textbf{Cadential Resolution:} Certain compositions generated by the system appear to lack a conclusive sense of resolution, resulting in pieces that may feel unfinished or conclude abruptly. This issue affects the listener's sense of closure and satisfaction, detracting from the overall effectiveness of the musical experience. The challenge in achieving cadential resolution is partly due to the inherent difficulty for GPTs to grasp the concept of musical closure, which the perpetual aspect of its nature is hard for a language model to handle.

\section{Conclusion}
In conclusion, our investigation into ComposerX, a multi-agent polyphonic symbolic music composition system, reveals its efficacy in leveraging the inherent musical capabilities of LLMs to compose high-quality music. By introducing a collaborative agent-based approach, ComposerX not only transcends the capabilities of single-agent systems but also offers a cost-effective alternative to traditional music generation models that rely heavily on computational resources.

\section{Contributions and Acknowledgments}
\textbf{Core}\\
Qixin Deng, \textit{qdeng4@u.rochester.edu}\\
Qikai Yang, \textit{qikaiy2@illinois.edu}\\
Ruibin Yuan, \textit{ryuanab@connect.ust.hk}\\
Yipeng Huang, \textit{hyp744009246@163.com}\\

\noindent\textbf{Contributors}\\
Yi Wang\\
Xubo Liu\\
Zeyue Tian\\
Jiahao Pan\\
Ge Zhang\\
Hanfeng Lin\\
Yizhi Li\\
Yinghao Ma\\
Jie Fu\\
Chenghua Lin\\
Emmanouil Benetos\\
Wenwu Wang\\
Gus Xia\\

\noindent\textbf{Correspondence}\\
Wei Xue, \textit{weixue@ust.hk}\\
Yike Guo, \textit{yikeguo@ust.hk}\\

\bibliography{ISMIRtemplate}

\begin{thebibliography}{10}
\providecommand{\url}[1]{#1}
\csname url@samestyle\endcsname
\providecommand{\newblock}{\relax}
\providecommand{\bibinfo}[2]{#2}
\providecommand{\BIBentrySTDinterwordspacing}{\spaceskip=0pt\relax}
\providecommand{\BIBentryALTinterwordstretchfactor}{4}
\providecommand{\BIBentryALTinterwordspacing}{\spaceskip=\fontdimen2\font plus
\BIBentryALTinterwordstretchfactor\fontdimen3\font minus \fontdimen4\font\relax}
\providecommand{\BIBforeignlanguage}[2]{{%
\expandafter\ifx\csname l@#1\endcsname\relax
\typeout{** WARNING: IEEEtran.bst: No hyphenation pattern has been}%
\typeout{** loaded for the language `#1'. Using the pattern for}%
\typeout{** the default language instead.}%
\else
\language=\csname l@#1\endcsname
\fi
#2}}
\providecommand{\BIBdecl}{\relax}
\BIBdecl

\bibitem{masataka2009origins}
N.~Masataka, ``The origins of language and the evolution of music: A comparative perspective,'' \emph{Physics of Life Reviews}, vol.~6, no.~1, pp. 11--22, 2009.

\bibitem{masataka2007music}
------, ``Music, evolution and language,'' \emph{Developmental science}, vol.~10, no.~1, pp. 35--39, 2007.

\bibitem{pino2023association}
M.~C. Pino, M.~Giancola, and S.~D’Amico, ``The association between music and language in children: A state-of-the-art review,'' \emph{Children}, vol.~10, no.~5, p. 801, 2023.

\bibitem{vaswani2017attention}
A.~Vaswani, N.~Shazeer, N.~Parmar, J.~Uszkoreit, L.~Jones, A.~N. Gomez, {\L}.~Kaiser, and I.~Polosukhin, ``Attention is all you need,'' \emph{Advances in neural information processing systems}, vol.~30, 2017.

\bibitem{huang2018musictransformer}
C.-Z.~A. Huang, A.~Vaswani, J.~Uszkoreit, N.~Shazeer, I.~Simon, C.~Hawthorne, A.~M. Dai, M.~D. Hoffman, M.~Dinculescu, and D.~Eck, ``Music transformer,'' \emph{arXiv preprint arXiv:1809.04281}, 2018.

\bibitem{payne2019musenet}
\BIBentryALTinterwordspacing
C.~Payne, ``Musenet,'' OpenAI Blog, Apr 2019. [Online]. Available: \url{https://openai.com/blog/musenet}
\BIBentrySTDinterwordspacing

\bibitem{lu2023musecoco}
P.~Lu, X.~Xu, C.~Kang, B.~Yu, C.~Xing, X.~Tan, and J.~Bian, ``Musecoco: Generating symbolic music from text,'' 2023.

\bibitem{dhariwal2020jukebox}
P.~Dhariwal, H.~Jun, C.~Payne, J.~W. Kim, A.~Radford, and I.~Sutskever, ``Jukebox: A generative model for music,'' \emph{arXiv preprint arXiv:2005.00341}, 2020.

\bibitem{agostinelli2023musiclm}
A.~Agostinelli, T.~I. Denk, Z.~Borsos, J.~Engel, M.~Verzetti, A.~Caillon, Q.~Huang, A.~Jansen, A.~Roberts, M.~Tagliasacchi \emph{et~al.}, ``Musiclm: Generating music from text,'' \emph{arXiv preprint arXiv:2301.11325}, 2023.

\bibitem{copet2023simple}
J.~Copet, F.~Kreuk, I.~Gat, T.~Remez, D.~Kant, G.~Synnaeve, Y.~Adi, and A.~D{\'e}fossez, ``Simple and controllable music generation,'' \emph{arXiv preprint arXiv:2306.05284}, 2023.

\bibitem{margulis2016repetition}
E.~H. Margulis and R.~Simchy-Gross, ``Repetition enhances the musicality of randomly generated tone sequences,'' \emph{Music Perception: An Interdisciplinary Journal}, vol.~33, no.~4, pp. 509--514, 2016.

\bibitem{dai2022missing}
S.~Dai, H.~Yu, and R.~B. Dannenberg, ``What is missing in deep music generation? a study of repetition and structure in popular music,'' \emph{arXiv preprint arXiv:2209.00182}, 2022.

\bibitem{jhamtani2019modeling}
H.~Jhamtani and T.~Berg-Kirkpatrick, ``Modeling self-repetition in music generation using generative adversarial networks,'' in \emph{Machine Learning for Music Discovery Workshop, ICML}, 2019.

\bibitem{qu2024mupt}
X.~Qu, Y.~Bai, Y.~Ma, Z.~Zhou, K.~M. Lo, J.~Liu, R.~Yuan, L.~Min, X.~Liu, T.~Zhang \emph{et~al.}, ``Mupt: A generative symbolic music pretrained transformer,'' \emph{arXiv preprint arXiv:2404.06393}, 2024.

\bibitem{yue2023mammoth}
X.~Yue, X.~Qu, G.~Zhang, Y.~Fu, W.~Huang, H.~Sun, Y.~Su, and W.~Chen, ``Mammoth: Building math generalist models through hybrid instruction tuning,'' \emph{arXiv preprint arXiv:2309.05653}, 2023.

\bibitem{roziere2023code}
B.~Roziere, J.~Gehring, F.~Gloeckle, S.~Sootla, I.~Gat, X.~E. Tan, Y.~Adi, J.~Liu, T.~Remez, J.~Rapin \emph{et~al.}, ``Code llama: Open foundation models for code,'' \emph{arXiv preprint arXiv:2308.12950}, 2023.

\bibitem{bubeck2023sparks}
S.~Bubeck, V.~Chandrasekaran, R.~Eldan, J.~Gehrke, E.~Horvitz, E.~Kamar, P.~Lee, Y.~T. Lee, Y.~Li, S.~Lundberg \emph{et~al.}, ``Sparks of artificial general intelligence: Early experiments with gpt-4,'' \emph{arXiv preprint arXiv:2303.12712}, 2023.

\bibitem{lin2024arrange}
L.~Lin, G.~Xia, Y.~Zhang, and J.~Jiang, ``Arrange, inpaint, and refine: Steerable long-term music audio generation and editing via content-based controls,'' \emph{arXiv preprint arXiv:2402.09508}, 2024.

\bibitem{lin2023content}
L.~Lin, G.~Xia, J.~Jiang, and Y.~Zhang, ``Content-based controls for music large language modeling,'' \emph{arXiv preprint arXiv:2310.17162}, 2023.

\bibitem{lin2023equipping}
------, ``Equipping musicgen with chord and rhythm controls,'' in \emph{Ismir 2023 Hybrid Conference}, 2023.

\bibitem{yuan2024chatmusician}
R.~Yuan, H.~Lin, Y.~Wang, Z.~Tian, S.~Wu, T.~Shen, G.~Zhang, Y.~Wu, C.~Liu, Z.~Zhou, Z.~Ma, L.~Xue, Z.~Wang, Q.~Liu, T.~Zheng, Y.~Li, Y.~Ma, Y.~Liang, X.~Chi, R.~Liu, Z.~Wang, P.~Li, J.~Wu, C.~Lin, Q.~Liu, T.~Jiang, W.~Huang, W.~Chen, E.~Benetos, J.~Fu, G.~Xia, R.~Dannenberg, W.~Xue, S.~Kang, and Y.~Guo, ``Chatmusician: Understanding and generating music intrinsically with llm,'' 2024.

\bibitem{ding2024songcomposer}
S.~Ding, Z.~Liu, X.~Dong, P.~Zhang, R.~Qian, C.~He, D.~Lin, and J.~Wang, ``Songcomposer: A large language model for lyric and melody composition in song generation,'' \emph{arXiv preprint arXiv:2402.17645}, 2024.

\bibitem{liang2024bytecomposer}
X.~Liang, J.~Lin, and X.~Du, ``Bytecomposer: a human-like melody composition method based on language model agent,'' \emph{arXiv preprint arXiv:2402.17785}, 2024.

\bibitem{wu2023exploring}
S.~Wu and M.~Sun, ``Exploring the efficacy of pre-trained checkpoints in text-to-music generation task,'' 2023.

\bibitem{zhang2023loop}
Y.~Zhang, A.~Maezawa, G.~Xia, K.~Yamamoto, and S.~Dixon, ``Loop copilot: Conducting ai ensembles for music generation and iterative editing,'' \emph{arXiv preprint arXiv:2310.12404}, 2023.

\bibitem{yu2023musicagent}
D.~Yu, K.~Song, P.~Lu, T.~He, X.~Tan, W.~Ye, S.~Zhang, and J.~Bian, ``Musicagent: An ai agent for music understanding and generation with large language models,'' \emph{arXiv preprint arXiv:2310.11954}, 2023.

\bibitem{wang2023selfinstruct}
Y.~Wang, Y.~Kordi, S.~Mishra, A.~Liu, N.~A. Smith, D.~Khashabi, and H.~Hajishirzi, ``Self-instruct: Aligning language models with self-generated instructions,'' 2023.

\bibitem{wei2023chainofthought}
J.~Wei, X.~Wang, D.~Schuurmans, M.~Bosma, B.~Ichter, F.~Xia, E.~Chi, Q.~Le, and D.~Zhou, ``Chain-of-thought prompting elicits reasoning in large language models,'' 2023.

\bibitem{wu2023autogen}
Q.~Wu, G.~Bansal, J.~Zhang, Y.~Wu, B.~Li, E.~Zhu, L.~Jiang, X.~Zhang, S.~Zhang, J.~Liu, A.~H. Awadallah, R.~W. White, D.~Burger, and C.~Wang, ``Autogen: Enabling next-gen llm applications via multi-agent conversation,'' 2023.

\bibitem{donahue2023singsong}
C.~Donahue, A.~Caillon, A.~Roberts, E.~Manilow, P.~Esling, A.~Agostinelli, M.~Verzetti, I.~Simon, O.~Pietquin, N.~Zeghidour \emph{et~al.}, ``Singsong: Generating musical accompaniments from singing,'' \emph{arXiv preprint arXiv:2301.12662}, 2023.

\end{thebibliography}

%
%
%
%
%
\clearpage 
\onecolumn 
\appendix

\end{document}